\documentclass[epjST]{svjour}
\usepackage{graphics}
\usepackage{amsmath}
\usepackage{amssymb}
\usepackage{float}
\usepackage{physics}
\usepackage{verbatim}
\usepackage{color}

\begin{document}
\title{Sparse-Hamiltonian approach to the time evolution of molecules on quantum computers}
\author{Christina Daniel\inst{1} \and Diksha Dhawan\inst{2} \and Dominika Zgid\inst{2,3} \and James K. Freericks\inst{1} }
\institute{Department of Physics, Georgetown University, Washington, DC 20057 USA \and Department of Chemistry, University of Michigan, Ann Arbor, Michigan 48109, USA \and Department of Physics, University of Michigan, Ann Arbor, Michigan 48109, USA}
\abstract{
Quantum chemistry has been viewed as one of the potential early applications of quantum computing. Two techniques have been proposed for electronic structure calculations: (i) the variational quantum eigensolver and (ii) the phase-estimation algorithm. In both cases, the complexity of the problem increases for basis sets where either the Hamiltonian is not sparse, or it is sparse, but many orbitals are required to accurately describe the molecule of interest. In this work, we explore the possibility of mapping the molecular problem onto a sparse Hubbard-like Hamiltonian, which allows a Green's-function-based approach to electronic structure via a hybrid quantum-classical algorithm. We illustrate the time-evolution aspect of this methodology with a simple four-site hydrogen ring. 
} 
\maketitle
\section{Introduction}
\label{intro}

In the variational quantum eigensolver algorithm~\cite{vqe,qc_chem_review}, one prepares a trial wavefunction and evaluates the expectation values needed to determine the expectation value of the Hamiltonian with respect to that wavefunction. The number of measurements scales with the number of nonzero terms in the Hamiltonian, which typically grows like the fourth power of the number of spin orbitals used in the basis set for the given calculation. The state is usually prepared with a simple strategy like a unitary coupled cluster approach~\cite{ucc}. A self-consistent loop optimizes the parameters in the variational ansatz until the required accuracy is achieved. The phase estimation algorithm~\cite{phase-estimation} instead determines the phase of $\exp ( -i\lambda \hat{\mathcal H} )$ and requires many operations of the exponential of the Hamiltonian onto the wavefunction, similar to time evolution, to complete the calculation ($\lambda$ is a scaling factor). If the initial wavefunction has high overlap with the ground state, then the chance to project onto the ground state with the measurement is high. 

In both cases, the complexity of the algorithm grows with the number of nonzero terms in the Hamiltonian matrix---for the variational quantum eigensolver, this is seen in the number of measurements required, while in the phase estimation algorithm it is in the number of independent Trotter steps required for each application of the exponential of the Hamiltonian (multiplied by a constant). Given the fact that current noisy intermediate scale quantum (NISQ) computers can only run low depth circuits, this is problematic for running these algorithms on complex molecules. Even when fault-tolerant quantum computers become available, they may still require low-depth circuits due to drift of the tuning of the machine over extended periods of time (which is not normally corrected by error correction algorithms). This then implies that methods focused on making the Hamiltonian matrix sparse are critical to the success of quantum chemistry applications on quantum computers in the near term. 

In this work, we describe the time-evolution piece of the algorithm to do this. It is based on a simple premise that the electron correlations in the molecule can be efficiently encoded in the self-energy of the molecule. Then, if we can construct a sparse Hamiltonian that approximates the self-energy of the molecule well, we can use it to determine the properties of the molecule. We describe just how such a process can be carried out on a quantum computer with a simple example below. We examine the accuracy of using an approximate unitary coupled cluster wavefunction to estimate the zero-temperature Green's function of the sparse Hamiltonian, which is the Hubbard Hamiltonian here.

\section{Formalism}

The retarded Green's function in position space is defined to be
\begin{equation}
    G_{ij\sigma}(t)=-i\theta(t)\mathrm{Tr}\, e^{-\beta\hat{\mathcal H}}\{\hat {c}_{i\sigma}^{\phantom\dagger}(t),\hat {c}_{j\sigma}^\dagger\}\frac{1}{\mathcal{Z}},
\end{equation}
where $\mathcal{Z}=\mathrm{Tr}\,\exp(-\beta\hat{\mathcal H})$ is the partition function, $\beta$ is the inverse temperature and $\theta(t)$ is the unit step function. Here we have that $\hat {c}_{i\sigma}^{\phantom\dagger}$ ($\hat{c}_{i\sigma}^\dagger$) are the annihilation (creation) operators for an electron at site $i$ with spin $\sigma$. The braces denote the anticommutator, and the time-evolution of the operators is given in the Heisenberg representation. The trace is over all many body states with a fixed number of electrons (that is, we are calculating a canonical, not a grand canonical Green's function here). In this work, we focus on $T=0$, where the trace includes just one state, the ground state. We also can work in momentum space (we assume the lattice is periodic), where 
\begin{equation}
    \hat{c}_{k\sigma}=\frac{1}{\sqrt{V}}\sum_{j=0}^{V-1}e^{-ikj}\hat{c}_{j\sigma},
\end{equation}
$V$ is the number of lattice sites and we set the lattice constant $a=1$. The allowed $k$ values are $0$, $\pi/2$, $\pi$, and $3\pi/2$ for a four-site lattice.

We will be mapping the hydrogen ring to a sparse Hamiltonian given by the Hubbard model~\cite{hubbard}, which is 
\begin{equation}
\hat{\mathcal{H}} = \sum_{ij\sigma} t_{ij} \hat{c}^{\dag}_{i\sigma}\hat{c}_{j\sigma}^{\phantom\dagger} 
+ U \sum_{i} \hat{c}^\dag_{i\uparrow}   \hat{c}_{i\uparrow}^{\phantom\dagger}   \hat{c}^\dag_{i\downarrow} \hat{c}_{i\downarrow}^{\phantom\dagger}.
\end{equation}
Here, $t_{ij}$ is the hopping matrix and $U$ is the on-site Coulomb interaction. The first term is  the kinetic energy, and the second term is the potential energy. In this mapping, the hopping matrix is a full matrix, with nonzero coefficients for all hopping terms.

The Hamiltonian can also be written in momentum space as
\begin{equation}
    \hat{\mathcal H}=\sum_{k,\sigma}\epsilon_{k}\hat{c}_{k\sigma}^\dagger\hat{c}_{k\sigma}^{\phantom\dagger}+\frac{U}{V}\sum_{kk'q}
    \hat{c}_{k\uparrow}^\dagger\hat{c}_{k'\uparrow}^{\phantom\dagger}\hat{c}_{q\downarrow}^\dagger\hat{c}_{k-k'+q\downarrow}^{\phantom\dagger}
\end{equation}
and we will be primarily working with this form. Here we have the bandstructure given by $\epsilon_{k}=\frac{1}{V}\sum_{jj'}t_{jj'}\exp(ikj)$ 
, which is independent of $j'$ due to the translational invariance of the lattice (hydrogen ring). If the molecule is not a translationally invariant ring, a more complicated single-particle term to the Hamiltonian is needed, but we do not discuss this further here. Note that the hopping matrix also includes diagonal terms with $j=j'$.

The mapping of the molecular Hamiltonian to the Hubbard Hamiltonian is designed to recover the dynamic part of the finite temperature self-energy of the parent molecular problem with all two-body interactions present. We call such a mapping the dynamical self-energy mapping (DSEM).
Such a mapping was described in Ref.~\cite{ZgidRusakov} where the effective on-site two-body integrals were chosen to recover the first moment of the frequency dependent self-energy. Here, as a proof of principle, the given sparse Hamiltonian is created to recover the first moment of the exact self-energy obtained in the exact-diagonalization procedure. In general, the following scheme of mapping can be used to design quantum--classical hybrid algorithms where a classical computer is used to calculate the sparse Hamiltonian that is then employed by the quantum computer. Details of the preparation of such a mapping are described in Ref.~\cite{this willcomethisweek}.

\section{Results}

The fitting procedure produces a diagonal term $t_0$, a nearest neighbor hopping $t_1$ and a second neighbor hopping $t_2$, along with the on-site repulsion $U$ 
 (see Table~\ref{tab:1}). 
\begin{table}[H]
\centering
\caption{Parameters for the sparse Hubbard Hamiltonian that represents the four-site hydrogen ring. ll parameters are in Hartrees.}
\label{tab:1}       
\begin{tabular}{llll}
\hline\noalign{\smallskip}
$U$ & $t_0$ & $t_1$ & $t_2$ \\
\noalign{\smallskip}\hline\noalign{\smallskip}
0.6830907036 & {-0.3025} & -0.380776 & 0.03035031  \\
\noalign{\smallskip}\hline
\end{tabular}
\end{table}

The exact ground state is found by diagonalizing the Hamiltonian with four electrons. It yields
\begin{equation}
    \begin{split}
\ket{\Psi_{0}}&=
\alpha 
\big(\hat{c}^\dag_{0\uparrow}\hat{c}^\dag_{1\uparrow}\hat{c}^\dag_{0\downarrow}\hat{c}^\dag_{1\downarrow}\ket{0} 
- \hat{c}^\dag_{0\uparrow}\hat{c}^\dag_{3\uparrow}\hat{c}^\dag_{0\downarrow}\hat{c}^\dag_{3\downarrow}\ket{0} \big)
+ \beta \big( 
\hat{c}^{\dag}_{0\uparrow} \hat{c}^{\dag}_{1\uparrow} \hat{c}^{\dag}_{2\downarrow} \hat{c}^{\dag}_{3\downarrow}\ket{0}
+ 
\hat{c}^\dag_{0\uparrow}\hat{c}^\dag_{3\uparrow}\hat{c}^\dag_{1\downarrow}\hat{c}^\dag_{2\downarrow}\ket{0} 
 \\
&+\hat{c}^\dag_{1\uparrow}\hat{c}^\dag_{2\uparrow}\hat{c}^\dag_{0\downarrow}\hat{c}^\dag_{3\downarrow}\ket{0} 
+  \hat{c}^\dag_{2\uparrow} \hat{c}^\dag_{3\uparrow} \hat{c}^\dag_{0\downarrow}\hat{c}^\dag_{1\downarrow}\ket{0} 
+  
2 \hat{c}^\dag_{1\uparrow}\hat{c}^\dag_{3\uparrow}\hat{c}^\dag_{0\downarrow}\hat{c}^\dag_{2\downarrow}\ket{0}
+ 2 \hat{c}^\dag_{0\uparrow}\hat{c}^\dag_{2\uparrow}\hat{c}^\dag_{1\downarrow}\hat{c}^\dag_{3\downarrow}\ket{0} \big)
\\ 
&+\gamma \big( \hat{c}^\dag_{2\uparrow} \hat{c}^\dag_{3\uparrow}\hat{c}^\dag_{2\downarrow}\hat{c}^\dag_{3\downarrow}\ket{0}
- \hat{c}^\dag_{1\uparrow} \hat{c}^\dag_{2\uparrow} \hat{c}^\dag_{1\downarrow}\hat{c}^\dag_{2\downarrow}\ket{0}  \big),
    \end{split}
\end{equation}
with
$\alpha=0.6895316741725$, $\beta=0.059610737681519$ and $\gamma=0.056792869544809$.
While we could compute the Green's function for the exact ground state, that would not be representative of what a true quantum computation would be that is based on the variational quantum eigensolver, so we instead use an approximate ground state based on a factorized unitary coupled cluster ansatz that uses just doubles excitations from the reference state where both the level with $k=0$ and the level with $k=1$ are filled. This state was described in Ref.~\cite{RefJ2}, where the excitation operators for a factorized unitary coupled cluster are given. That approach is generalized here and summarized in Tables \ref{tab:2} and \ref{tab:3}. 

The factorized form of the unitary coupled cluster approximation applies each doubles excitation (and de-excitation) operator in the order given in Table~\ref{tab:2} to the initial reference state, $\hat{c}^\dag_{0\uparrow}\hat{c}^\dag_{1\uparrow}\hat{c}^\dag_{0\downarrow}\hat{c}^\dag_{1\downarrow}\ket{0} $. The resulting, approximate ground state in terms of the three angles $\theta_1$, $\theta_3$, and $\theta_4$ is summarized in Table~\ref{tab:3}. We use the same notation as used in Ref.~\cite{RefJ2}, which is why we have no $\theta_2$, since that was used for a quad excitation that we do not include here.

\begin{table}[htb]
\centering
\caption{Doubles unitary coupled-cluster operators used in creating the approximate ground state. The operators are applied in order according to the rows of the table.}
\label{tab:2}       
\begin{tabular}{c | l}
\hline\noalign{\smallskip}
Order& Unitary coupled cluster factor \\
\noalign{\smallskip}\hline\noalign{\smallskip}
1&$e^{-\theta_1  \hat{c}_{2\uparrow}^\dag \hat{c}^\dag_{3\downarrow} \hat{c}_{0\downarrow}^{\phantom\dag}\hat{c}_{1\uparrow}^{\phantom\dag} + \theta_1 \hat{c}^\dag_{1\uparrow} \hat{c}^\dag_{0\downarrow} \hat{c}_{3\downarrow}^{\phantom\dag} \hat{c}_{2\uparrow}^{\phantom\dag} }$ \\
2&$e^{-\theta_1 \hat{c}_{3\uparrow}^\dag \hat{c}_{2\downarrow}^\dag\hat{c}_{1\downarrow}^{\phantom\dag}\hat{c}_{0\uparrow}^{\phantom\dag} + \theta_1 \hat{c}^\dag_{0\uparrow} \hat{c}^\dag_{1\downarrow} \hat{c}_{2\downarrow}^{\phantom\dag} \hat{c}_{3\uparrow}^{\phantom\dag}}$  \\ 
3&$e^{-\frac{\pi}{4} \hat{c}_{3\uparrow}^\dag\hat{c}_{3\downarrow}^\dag\hat{c}_{1\downarrow}^{\phantom\dag}\hat{c}_{1\uparrow}^{\phantom\dag} + \frac{\pi}{4} \hat{c}^\dag_{1\uparrow} \hat{c}^\dag_{1\downarrow} \hat{c}_{3\downarrow}^{\phantom\dag} \hat{c}_{3\uparrow}^{\phantom\dag}}$  \\ 
4&$e^{+\theta_3 \hat{c}_{2\uparrow}^\dag\hat{c}_{3\uparrow}^\dag\hat{c}_{1\uparrow}^{\phantom\dag}\hat{c}_{0\uparrow}^{\phantom\dag} - \theta_3 \hat{c}^\dag_{0\uparrow} \hat{c}^\dag_{1\uparrow} \hat{c}_{3\uparrow}^{\phantom\dag} \hat{c}_{2\uparrow}^{\phantom\dag}}$  \\ 
5&$e^{+\theta_3 \hat{c}_{2\downarrow}^\dag\hat{c}_{3\downarrow}^\dag\hat{c}_{1\downarrow}^{\phantom\dag}\hat{c}_{0\downarrow}^{\phantom\dag} - \theta_3 \hat{c}^\dag_{0\downarrow} \hat{c}^\dag_{1\downarrow} \hat{c}_{3\downarrow}^{\phantom\dag} \hat{c}_{2\downarrow}^{\phantom\dag} }$ \\ 
6&$e^{-\theta_3 \hat{c}_{1\uparrow}^\dag\hat{c}_{2\uparrow}^\dag\hat{c}_{3\uparrow}^{\phantom\dag}\hat{c}_{0\uparrow}^{\phantom\dag} + \theta_3 \hat{c}^\dag_{0\uparrow} \hat{c}^\dag_{3\uparrow} \hat{c}_{2\uparrow}^{\phantom\dag} \hat{c}_{1\uparrow}^{\phantom\dag}} $ \\ 
7&$e^{-\theta_3 \hat{c}_{1\downarrow}^\dag\hat{c}_{2\downarrow}^\dag\hat{c}_{3\downarrow}^{\phantom\dag}\hat{c}_{0\downarrow}^{\phantom\dag} + \theta_3 \hat{c}^\dag_{0\downarrow} \hat{c}^\dag_{3\downarrow} \hat{c}_{2\downarrow}^{\phantom\dag} \hat{c}_{1\downarrow}^{\phantom\dag}}$  \\ 
8&$e^{-\theta_4 \hat{c}_{2\uparrow}^\dag\hat{c}_{2\downarrow}^\dag\hat{c}_{0\downarrow}^{\phantom\dag}\hat{c}_{0\uparrow}^{\phantom\dag} + \theta_4 \hat{c}^\dag_{0\uparrow} \hat{c}^\dag_{0\downarrow} \hat{c}_{2\downarrow}^{\phantom\dag} \hat{c}_{2\uparrow}^{\phantom\dag}}$ \\
\noalign{\smallskip}\hline
\end{tabular}
\end{table}
Once the analytical coefficients of Table~\ref{tab:3} are obtained, numerical values are calculated from the analytical coefficients. To do this, equations for the angles $\theta_1$, $\theta_3$, and $\theta_4$ are needed.  These equations are given in Ref. \cite{RefJ2} and the angles depend on the values of $\alpha$, $\beta$, and $\gamma$ from the exact ground state. They are 
\begin{align}
    \theta_1 &= \frac{1}{2}\sin^{-1}(4\beta) \\
    \theta_3 &= \frac{1}{2}\sin^{-1}\left(\frac{2\sqrt{2}\beta}{c^2_1}\right) \\
    \theta_4 &= \tan^{-1}\left(\frac{\gamma}{\alpha}\right) - \tan^{-1}\left (\tan^2\theta_3\right ).
\end{align}
After substituting the values for $\alpha$, $\beta$, and $\gamma$ into the equations for the angles, the approximate ground state with numerical coefficients is obtained (see  Table~\ref{tab:3}). This wavefunction is representative of a generic state that one would obtain after performing a variational quantum eigensolver calculation.

\begin{table}[htb]
\centering
\caption{General form and final numerical values of the coefficients of the approximate ground state after applying the doubles-only excitations, with $c_i \equiv \cos{\theta_i}$ and $s_i \equiv \sin{\theta_i}$.}
\label{tab:3}       
\begin{tabular}{l l l}
\hline\noalign{\smallskip}
State & Analytical Coefficient & Numerical Coefficient\\
\noalign{\smallskip}\hline\noalign{\smallskip}
$\hat{c}^\dag_{0\uparrow}\hat{c}^\dag_{1\uparrow}\hat{c}^\dag_{0\downarrow}\hat{c}^\dag_{1\downarrow}\ket{0} $ &  $\frac{c_4}{\sqrt{2}} ({c_1^2} c_3^2 + {s_1^2} s_3^2) + \frac{s_4}{\sqrt{2}} ({s_1^2} c_3^2 -{c_1^2} s_3^2)$  & 0.6902877166375496 \\
$\hat{c}^\dag_{0\uparrow}\hat{c}^\dag_{3\uparrow}\hat{c}^\dag_{0\downarrow}\hat{c}^\dag_{3\downarrow}\ket{0}$ & $\frac{c_4}{\sqrt{2}} ({s_1^2} s_3^2 -{c_1^2} c_3^2 ) + \frac{s_4}{\sqrt{2}}({s_1^2} c_3^2 + {c_1^2} s_3^2) $ & -0.6886258223794277\\
$\hat{c}^{\dag}_{0\uparrow} \hat{c}^{\dag}_{1\uparrow} \hat{c}^{\dag}_{2\downarrow} \hat{c}^{\dag}_{3\downarrow}\ket{0}$ & $ \frac{c_1^2}{\sqrt{2}} s_3 c_3 - \frac{s_1^2}{\sqrt{2}} s_3 c_3$ & 0.05873846703927717 \\
$\hat{c}^\dag_{0\uparrow}\hat{c}^\dag_{3\uparrow}\hat{c}^\dag_{1\downarrow}\hat{c}^\dag_{2\downarrow}\ket{0}$ & $\frac{c_3 s_3}{\sqrt{2}}$ & 0.06048300832376081\\
$\hat{c}^\dag_{1\uparrow}\hat{c}^\dag_{2\uparrow}\hat{c}^\dag_{0\downarrow}\hat{c}^\dag_{3\downarrow}\ket{0} $ & $\frac{c_3 s_3}{\sqrt{2}}$ & 0.06048300832376081\\
$\hat{c}^\dag_{2\uparrow} \hat{c}^\dag_{3\uparrow} \hat{c}^\dag_{0\downarrow}\hat{c}^\dag_{1\downarrow}\ket{0} $ & $\frac{c_1^2}{\sqrt{2}} s_3 c_3 - \frac{s_1^2}{\sqrt{2}} s_3 c_3 $ & 0.05873846703927717 \\
$\hat{c}^\dag_{1\uparrow}\hat{c}^\dag_{3\uparrow}\hat{c}^\dag_{0\downarrow}\hat{c}^\dag_{2\downarrow}\ket{0}$ & $c_1 s_1$  & 0.11922147536303802\\
$ \hat{c}^\dag_{0\uparrow}\hat{c}^\dag_{2\uparrow}\hat{c}^\dag_{1\downarrow}\hat{c}^\dag_{3\downarrow}\ket{0}$ & $c_1 s_1$  & 0.11922147536303802 \\
$\hat{c}^\dag_{2\uparrow} \hat{c}^\dag_{3\uparrow}\hat{c}^\dag_{2\downarrow}\hat{c}^\dag_{3\downarrow}\ket{0}$ & $\frac{c_4}{\sqrt{2}} ({s_1^2} c_3^2 + {c_1^2} s_3^2)  + \frac{s_4}{\sqrt{2}} ({c_1^2} c_3^2 - {s_1^2} s_3^2) $ & 0.06687536735226934\\ 
$\hat{c}^\dag_{1\uparrow} \hat{c}^\dag_{2\uparrow} \hat{c}^\dag_{1\downarrow}\hat{c}^\dag_{2\downarrow}\ket{0}$ & $ \frac{c_4}{\sqrt{2}}({s_1^2}c_3^2 -{c_1^2} s_3^2) - \frac{s_4}{\sqrt{2}}({c_1^2} c_3^2 + {s_1^2} s_3^2)$ & -0.04669803278042805\\
\noalign{\smallskip}\hline
\end{tabular}
\end{table}

\begin{figure}[H]
\centering
\resizebox{1\columnwidth}{!}
{\includegraphics{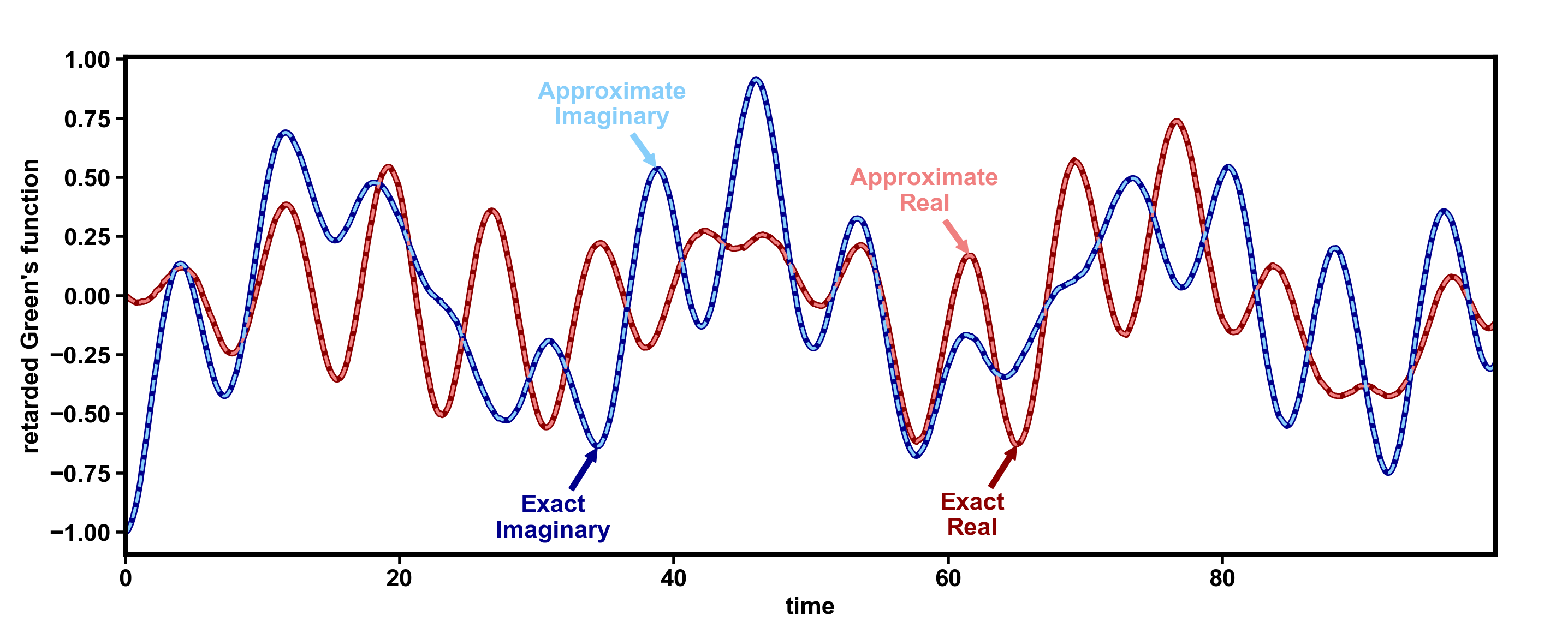}}
\caption{Exact (solid) and approximate (dashed) retarded Green's function for the sparse Hubbard Hamiltonian. The time is in units of $\hbar/$H.}
\label{fig:1}
\end{figure}

Using this approximate ground state, we compute the (approximate) time-dependent Green's function, where the ground-state wavefunction is replaced by the approximate ground-state, and compare it to the exact Green's function with the exact ground state. Figure \ref{fig:1} shows these two results (real and imaginary part). One can see that the two Green's functions are nearly identical; their differences are on the order of $10^{-4}$. The reason why is that the square of the overlap of the approximate state with the ground state is very high (the fidelity is 0.99979). This value is surprising given that the approximate ground state differs from the exact one with coefficients that are on the order of 0.01, but there is a cancellation leading to a higher fidelity.

Unfortunately, for such a small system, it does not make sense to use Dyson's equation to extract the frequency-dependent self-energy, because the data is truncated to too short of a time. This leads to inaccuracies here, because the system is finite and the Fourier transform of the Green's function yields a sum over delta functions. But if the Green's function is cut-off too early in the time domain, the results of the Fourier transform would be significantly distorted. This becomes less of a concern for larger molecules because they have so many frequencies that they tend to dephase and create a decaying Green's function in the time domain, which can be cut off, resulting in just a small broadening of the delta functions, so the calculations can proceed more normally for larger systems. Most molecules will fall into this category.

\section{Discussion}
\label{sec:7}

The algorithm on the quantum computer is now straightforward. After mapping the problem from the molecule to the sparse Hubbard Hamiltonian, we use the factorized form of the unitary coupled cluster ansatz to create an approximate ground state. In general, such an approach will involve many different types of excitations, but in this simple example, it involves only doubles excitations. The lower the order of the excitation, the lower the depth of the circuit for the quantum computer, so it is likely that many calculations will opt to only use singles and doubles, if possible. Then we would invoke the algorithm from the Los Alamos group~\cite{ortiz,troyer} to calculate the Green's function by measuring the $x$ or $y$ Pauli spin operator on the ancilla qubit to determine the real and imaginary parts of $G$. Both of these steps can be carried out with relatively low-depth circuits due to the sparsity of the effective Hamiltonian, but they do still require time evolution, which is still beyond the capability of currently available NISQ machines. Once one has determined the Green's function to far enough time on the quantum computer, then we would take the Fourier transform, extract the self-energy from Dyson's equation, and employ it to describe the Green's function of the molecule. This then allows the ground-state energy of the molecule to be determined.

\section{Conclusions}

In this work, we described an approach to use on near-term quantum computers that will allow us to calculate the electronic structure of more complex molecules sooner. The approach maps the molecule onto a sparse Hamiltonian that has a full single-particle hopping matrix, but only local interactions. Due to the significant reduction in the number of nonzero matrix elements, this sparse Hamiltonian becomes much easier to simulate on a quantum computer and one should be able to determine it's Green's function once time evolution becomes possible; this may occur in near-term NISQ machines or may need to wait until fault-tolerant computers are available. Once the Green's function for the sparse Hamiltonian is found on the quantum computer, we extract the self-energy and use it as the self-energy for the molecule in the full molecular problem. We showed how using an approximate form for the ground state leads to an accurate approximation to the exact result for a relatively long period of time. Hence, this makes it promising that such an approach can lead to accurate and efficient ways to perform electronic structure calculations on quantum computers. 

In terms of quantum computing complexity, a factorized form of UCC state preparation uses only doubles excitations here, and the time-evolution needed to compute the Green's function requires only control operations for the application of the creation or annihilation operators, but not for the time evolution. This is contrary to the quantum phase estimation algorithm, which requires controlled application of the Hamiltonian. Hence, the approach described here has the potential to be quite efficient for implementation on both near-term and fault-tolerant quantum computers.

\begin{acknowledgement}
This work is supported from the National Science Foundation under grant number CHE-1836497. JKF is also funded by the McDevitt bequest at Georgetown University. We thank Manuel Weber for useful discussions.
\end{acknowledgement}

\end{document}